\documentclass[aps,nofootinbib,showpacs]{revtex4}
\usepackage{graphicx}
\usepackage{amsfonts}
\usepackage{amsmath}

\usepackage{color}

\begin{document}

\title{Pair Production in Near Extremal Kerr-Newman Black Holes}

\author{Chiang-Mei Chen} \email{cmchen@phy.ncu.edu.tw}
\affiliation{Department of Physics, National Central University, Chungli 32001, Taiwan}
\affiliation{Center for Mathematics and Theoretical Physics, National Central University, Chungli 32054, Taiwan}

\author{Sang Pyo Kim} \email{sangkim@kunsan.ac.kr}
\affiliation{Department of Physics, Kunsan National University, Kunsan 54150, Korea}

\author{Jia-Rui Sun} \email{sunjiarui@sysu.edu.cn}
\affiliation{School of Physics and Astronomy, Sun Yat-Sen University, Guangzhou 510275, China}

\author{Fu-Yi Tang} \email{foue.tang@gmail.com}
\affiliation{Department of Physics, National Central University, Chungli 32001, Taiwan}

\date{\today}

%% REVTEX4
%\maketitle

\begin{abstract}
The spontaneous pair production of charged scalars in a near extremal Kerr-Newman (KN) black hole is analytically studied. It is shown that the existence condition for the pair production is equivalent to the violation of the Breitenlohner-Freedman bound in an AdS$_2$ space. The mean number of produced pairs in the extremal black hole has a thermal interpretation, in which the effective temperature for the Schwinger effect in the AdS$_2$ space persistently holds, while the mean number in the near extremal black hole has an additional factor of the Schwinger effect in the Rindler space. In addition, the holographic dual conformal field theory (CFT) descriptions of the charged scalar pair production are respectively realized both in the $J$ and $Q$ pictures in terms of the KN/CFTs correspondence.
\end{abstract}

%% REVTEX4

\maketitle
%\tableofcontents

%%%%%%%%%%%%%%%%%%%%%%%%%%%%%%%%%%%%%%%%%%%%%%%%%%%%%%%%%%%%%%%%%%%%%%
\section{Introduction}
%%%%%%%%%%%%%%%%%%%%%%%%%%%%%%%%%%%%%%%%%%%%%%%%%%%%%%%%%%%%%%%%%%%%%%
The spontaneous pair production from a black hole provides important information about quantum aspects of the vacuum in the curved spacetime. Especially, pair production from a charged black hole intertwines two remarkable quantum processes, namely the Schwinger mechanism~\cite{Schwinger:1951nm} caused by electromagnetic force and the Hawking radiation by tunneling through horizon~\cite{Parikh:1999mf}. The leading term of the pair-production rate is expected to come from the associated processes occurring at the near horizon region: the separation of virtual pairs by the causal boundary for the Hawking radiation and by the dominating electric field for the Schwinger mechanism. Therefore, one can mainly focus on quantum field theory in the near horizon geometry of charged black holes and obtain the pair-production rate. In particular, in the near extremal limit, the near horizon geometry turns out to have an enhanced symmetry. A well-known example is the near horizon geometry of a near extremal Reissner-Nordstr\"om (RN) black hole, which is the Bertotti-Robinson solution with the ${\rm AdS}_2 \times S^2$ isometry. The existence of such AdS structures in the background spacetime allows one to obtain analytical results for the field equation.

In the previous study~\cite{Chen:2012zn, Chen:2014yfa, Chen:2015fkx}, the spontaneous pair production of charged scalars and spinors has been systematically investigated in the near extremal RN black hole without back reactions. Interestingly, it has been shown that the existence condition for the pair production is equivalent to the violation of the Breitenlohner-Freedman (BF) bound in the AdS$_2$ (or warped AdS$_3$) spacetime, which, in turn, just guarantees the cosmic censorship conjecture during the pair production process. From the dual conformal field theory (CFT) side, the violation of the BF bound for the probe field makes the conformal dimensions of its dual operator complex numbers, which indicates the instability for the dual CFT. Following the techniques in the RN/CFT correspondence~\cite{Chen:2009ht, Chen:2010bsa, Chen:2010as, Chen:2011gz}, the holographic description of the pair production in the near extremal RN black hole has been found.

The pair production from a near extremal RN black hole exhibits interesting features~\cite{Chen:2012zn, Chen:2014yfa}, in particular, a thermal interpretation. The ${\rm AdS}_2 \times S^2$ geometry in the near horizon region of the extremal RN black hole gives, except for different quantum numbers for $S^2$, the same Schwinger effect in a constant electric field in the ${\rm AdS}_2$ space, which has a thermal interpretation in terms of the effective temperature consisting of the Unruh temperature for accelerating charges due to the electric field and the ${\rm AdS}$ curvature~\cite{Cai2014}. The effective temperature is an extension of the Unruh temperature in ${\rm AdS}_2$~\cite{Deser1997} to the Schwinger effect in that space. The pair-production rate in the near extremal RN black hole, however, has an additional factor due to a small deviation from the extremality, which has also another interpretation of the Schwinger effect in a Rindler space with the surface gravity of black hole for the acceleration~\cite{Kim2015a}. The Schwinger effect and the thermal interpretation in Ref.~\cite{Cai2014} seems to persistently hold as far as pairs are dominantly produced in the region of ${\rm AdS}_2 \times {\cal C}$ for a compact space ${\cal C}$.

Another interesting example is the near horizon geometry of a near extremal Kerr-Newman (KN) black hole which contains a warped $\mathrm{AdS}_3 = \mathrm{AdS}_2 \times S^1$ structure. It is thus expected to obtain analytical results for the pair production of probe field in that geometry. In addition, as has been shown, the RN/CFT duality can be incorporated into the $Q$ (charge) picture of the more general KN/CFTs dualities~\cite{Hartman:2008pb, Hartman:2009nz, Rasmussen:2010xd, Chen:2010ywa, Chen:2012np, Chen:2010jj}, while there exists another $J$ (angular momentum) dual CFT picture in terms of the Kerr/CFT duality~\cite{Guica:2008mu, Bredberg:2009pv, Compere:2012jk}. Hence, it is interesting to check the holographic description of the pair production in the near extremal KN black hole in both of the $Q$ and the $J$ pictures.

In this paper, we will study the spontaneous pair production process of charged scalars in the near extremal KN black hole without back reactions. The Klein-Gorden equation of the probe charged scalar field can be separated into the radial and angular equations. Though the angular coordinates are not spherical symmetric and the separation constant can only be numerically determined, the angular part is expected to contribute the same factor to the flux at the near horizon region and at the asymptotic AdS boundary. Therefore, the angular part will not affect the vacuum persistence amplitude, the mean number of produced pairs and the absorption cross section ratio. The radial equation, our main consideration, can be solved exactly in terms of the hypergeometric functions. Following previous works~\cite{Chen:2012zn, Chen:2014yfa}, we adopt the ``particle viewpoint'' and impose the boundary condition of no incoming flux at the asymptotic boundary. Using the fluxes, we then derive the Bogoliubov coefficients (vacuum persistence and mean number of pairs) and the absorption cross section ratio.

We also propose a thermal interpretation for the pair production in the extremal KN black hole, in which the effective temperature of the Unruh temperature for accelerating charges and the ${\rm AdS}$ curvature exactly gives the same form as that in the extremal RN black hole. The thermal interpretation holds for the near extremal KN black hole as for the near extremal RN black hole. To explain a physical origin of the leading term for the pair production, we employ the Hamilton-Jacobi action approach and then compute the instanton action from the phase-integral formula in the complex plane of the near horizon region. The Boltzmann factor for the Schwinger effect is a consequence of two simple poles at both the inner and outer horizons, in which there is no reason to exclude the quantum tunneling process from the inner horizon just adjacent to the outer horizon of the near extremal black hole. We show that the Unruh temperature in the effective temperature is the acceleration of charges due to the electric field on the polar points. Finally, based on the near-extremal KN/CFTs correspondence~\cite{Hartman:2008pb, Hartman:2009nz, Rasmussen:2010xd, Chen:2010ywa, Chen:2012np, Chen:2010jj}, we investigate the holographic dual CFTs descriptions of the scalar pair production in both of the $J$ and $Q$ pictures by comparing of the absorption cross section ratio computed both from the near extremal KN black hole and their dual CFTs, and show that they agree with each other, namely, the pair production in near extremal KN black hole indeed can be characterized by two individual CFTs by taking appropriate limits.

The outline of the paper is as follows. In Sec.~\ref{nhgKN} we review basic properties of the near horizon geometry of the near extremal KN black hole. In Secs.~\ref{scalarKN} and~\ref{pairproduction}, we analytically solve the field equation for the probe charged scalar field in the near extremal KN black hole and obtain the vacuum persistence amplitude, the mean number of produced pairs and the absorption cross section ratio. In Sec.~\ref{thermalinter}, the thermal interpretation of the pair production is presented. Then, in Sec.~\ref{cftdescription} we give the holographic descriptions of the pair production based on the KN/CFTs dualities. The conclusion is drawn in Sec.~\ref{conclusion} and in Appendix~\ref{AHF} we list some useful properties of the hypergeometric functions.

%%%%%%%%%%%%%%%%%%%%%%%%%%%%%%%%%%%%%%%%%%%%%%%%%%%%%%%%%%%%%%%%%%%%%%
\section{Near horizon geometry of Kerr-Newman}\label{nhgKN}
%%%%%%%%%%%%%%%%%%%%%%%%%%%%%%%%%%%%%%%%%%%%%%%%%%%%%%%%%%%%%%%%%%%%%%
The four-dimensional Einstein-Maxwell theory
\begin{equation}
S = \frac1{16 \pi} \int d^4x \sqrt{-g} \left( R - F_{[2]}^2 \right),
\end{equation}
admits the Kerr-Newman (KN) black hole solution, in natural unit: $c = G = \hbar = 1$, as follows
\begin{eqnarray} \label{KN}
ds^2 &=& - \frac{\Delta - a^2 \sin^2\theta}{\Sigma} \left[ d\hat{t} + \frac{(2 M \hat{r} - Q^2) a \sin^2\theta}{\Delta - a^2 \sin^2\theta} d\hat{\varphi} \right]^2 + \Sigma \left( \frac{d\hat{r}^2}{\Delta} + d\theta^2 + \frac{\Delta \sin^2\theta}{\Delta - a^2\sin^2\theta} d\hat{\varphi}^2 \right),
\nonumber\\
A_{[1]} &=& \frac{Q \hat{r}}{\Sigma} \left( d\hat{t} - a \sin^2\theta d\hat{\varphi} \right),
\end{eqnarray}
where
\begin{equation}
\Sigma = \hat{r}^2 + a^2 \cos^2\theta, \qquad \Delta = \hat{r}^2 - 2 M \hat{r} + a^2 + Q^2.
\end{equation}
This is the most general black hole carrying three physical quantities: mass $M$, electric charge $Q$, and angular momentum $J = M a$. The inner horizon $\hat{r}_-$ and the outer horizon $\hat{r}_+$ are located at $\hat{r}_\pm = M \pm \sqrt{M^2 - a^2 - Q^2}$. The thermodynamical properties of KN black holes are described by two essential quantities, the Hawking temperature and the Bekenstein-Hawking entropy
\begin{equation} \label{BHentropy}
T_\mathrm{H} = \frac{\hat{\kappa}}{2 \pi} = \frac{\hat{r}_+ - \hat{r}_-}{4 \pi (\hat{r}_+^2 + a^2)}, \qquad S_\mathrm{BH} = \frac{\hat{A}_+}{4} = \pi (\hat{r}_+^2 + a^2),
\end{equation}
where $\hat{\kappa}$ and $\hat{A}_+$ are the surface gravity and the area of the outer horizon. Beside, the first law of thermodynamics of the KN black hole reads
\begin{eqnarray}
\delta M = T_\mathrm{H} \delta S_\mathrm{BH} + \Phi_\mathrm{H} \delta Q + \Omega_\mathrm{H} \delta J,
\end{eqnarray}
in which $\Phi_\mathrm{H} = \frac{Q \hat{r}_+}{\hat{r}_+^2 + a^2}$ is the electric potential at the polar points of the horizon and $\Omega_\mathrm{H} = \frac{a}{\hat{r}_+^2 + a^2}$ is the angular velocity of the KN black hole.

From the ``orthodox'' limit for rotating back holes~\cite{Bardeen:1999px}, it is convenient to rewrite the KN metric in the ADM form as
\begin{eqnarray}
ds^2 &=& - \frac{\Sigma \Delta}{(\hat{r}^2 + a^2)^2 - \Delta a^2 \sin^2\theta} d\hat{t}^2 + \frac{\Sigma}{\Delta} d\hat{r}^2 + \Sigma d\theta^2
\nonumber\\
&& + \frac{(\hat{r}^2 + a^2)^2 - \Delta a^2 \sin^2\theta}{\Sigma} \sin^2\theta \left[ d\hat{\varphi} - \frac{a (2 M \hat{r} - Q^2)}{(\hat{r}^2 + a^2)^2 - \Delta a^2 \sin^2\theta} d\hat{t} \right]^2.
\end{eqnarray}
In the extremal limit, i.e. $M^2 = r_0^2$ with $r_0^2 \equiv Q^2 + a^2$, the horizons degenerate to $\hat{r} = \hat{r}_+ = \hat{r}_- = r_0$. To derive the near horizon geometry of a near extremal KN black hole, the angular velocity $\Omega_\mathrm{H}$ at the horizon should be firstly ``removed'' by the coordinate transformation
\begin{equation} \label{phicoordi}
\hat{\varphi} \to \varphi - \frac{a}{r_0^2 + a^2} \hat{t},
\end{equation}
and then taking the following near horizon and near extremal limit with $\varepsilon \to 0$,
\begin{equation} \label{NHKNlimit}
\hat{r} \to r_0 + \varepsilon r, \qquad \hat{t} \to \frac{r_0^2 + a^2}{\varepsilon} t, \qquad M \to r_0 + \varepsilon^2 \frac{B^2}{2 r_0},
\end{equation}
one obtains the near horizon solution of the near extremal KN black hole as
\begin{eqnarray} \label{NHKN}
ds^2 &=& \Gamma(\theta) \left[ -(r^2 - B^2) dt^2 + \frac{dr^2}{r^2 - B^2} + d\theta^2 \right] + \gamma(\theta) (d\varphi + b r dt)^2,
\\
A_{[1]} &=& - Q \left( \frac{r_0^2 - a^2 \cos^2\theta}{r_0^2 + a^2 \cos^2\theta} r dt + \frac{r_0 a \sin^2\theta}{r_0^2 + a^2 \cos^2\theta} d\varphi \right),
\end{eqnarray}
where
\begin{equation}
\Gamma(\theta) = r_0^2 + a^2 \cos^2\theta, \qquad \gamma(\theta) = \frac{(r_0^2 + a^2)^2 \sin^2\theta}{r^2_0 + a^2 \cos^2\theta}, \qquad b = \frac{2 a r_0}{r_0^2 + a^2}.
\end{equation}
The spacetime~(\ref{NHKN}) contains a warped AdS$_3$ geometry, which allows the dual CFTs description for the KN black hole.

%%%%%%%%%%%%%%%%%%%%%%%%%%%%%%%%%%%%%%%%%%%%%%%%%%%%%%%%%%%%%%%%%%%%%%
\section{Scalar Field in KN black hole}\label{scalarKN}
%%%%%%%%%%%%%%%%%%%%%%%%%%%%%%%%%%%%%%%%%%%%%%%%%%%%%%%%%%%%%%%%%%%%%%
The action for a probe charged scalar field $\Phi$ with the mass $m$ and the charge $q$ is given by
\begin{equation} \label{action}
S = \int d^4x \sqrt{-g} \left( - \frac12 D_\alpha \Phi^* D^\alpha \Phi - \frac12 m^2 \Phi^*\Phi \right),
\end{equation}
where $D_\alpha \equiv \nabla_\alpha - i q A_\alpha$ with $\nabla_\alpha$ being the covariant derivative in a curved spacetime. The corresponding Klein-Gordon equation is
\begin{equation} \label{eom}
(\nabla_\alpha - i q A_\alpha) (\nabla^\alpha - i q A^\alpha) \Phi - m^2 \Phi = 0.
\end{equation}
The scalar field in the metric (\ref{NHKN}) is assumed to have the following form
\begin{equation} \label{ansatz}
\Phi(t, r, \theta, \varphi) = \mathrm{e}^{-i \omega t + i n \varphi} R(r) \Theta(\theta),
\end{equation}
which separates the Klein-Gordon equation as
\begin{eqnarray}\label{radialeq}
\partial_r \left[ (r^2 - B^2) \partial_r R \right] + \left( \frac{\left[ \omega (r_0^2 + a^2) - q Q^3 r + 2 n a r_0 r \right]^2}{(r_0^2 + a^2)^2 (r^2 - B^2)} - m^2 (r_0^2 + a^2) - \lambda \right) R &=& 0, \label{EqR}
\\
\frac1{\sin\theta} \partial_\theta (\sin\theta \partial_\theta \Theta) - \left( \frac{\left[ n (r_0^2 + a^2 \cos^2\theta) + q Q a r_0 \sin^2\theta \right]^2}{(r_0^2 + a^2)^2 \sin^2\theta} - m^2 a^2 \sin^2\theta - \lambda \right) \Theta &=& 0,
\end{eqnarray}
where $\lambda$ is a separation constant. In addition, the radial equation~(\ref{radialeq}) may be viewed as the equation of motion for a probe scalar field with the effective mass $m^2_\mathrm{eff} = m^2 - \frac{(2 n a r_0 - q Q^3)^2}{(r_0^2 + a^2)^3} + \frac{\lambda}{r_0^2 + a^2}$ propagating in an AdS$_2$ spacetime with the AdS radius $L = \sqrt{r_0^2 + a^2}$. It has been shown from the pair production of scalars and spinors in the near extremal RN black hole that the existence condition for pair production is the appearance of an instability, i.e., the violation of the Breitenlohner-Freedman (BF) bound of the field in the AdS$_2$ spacet, which guarantees the existence of the propagating modes of the field~\cite{Chen:2012zn, Chen:2014yfa}. For the near extremal KN black hole, the corresponding relation is
\begin{eqnarray}\label{BFbv}
m^2_\mathrm{eff} < - \frac{1}{4 (r_0^2 + a^2)} \quad \Rightarrow \quad m^2 (r_0^2 + a^2) - \frac{(2 n a r_0 - q Q^3)^2}{(r_0^2 + a^2)^2} + \lambda + \frac14 < 0.
\end{eqnarray}

The radial flux of scalar field~(\ref{ansatz}) in the spacetime~(\ref{NHKN}) can be expressed as
\begin{eqnarray} \label{fluxD}
D &=& \int d\theta d\varphi i \sqrt{-g} g^{rr} (\Phi D_r \Phi^* - \Phi^* D_r \Phi)
\nonumber\\
&=& i (r_0^2 + a^2) (r^2 - B^2) (R \partial_r R^* - R^* \partial_r R) \mathfrak{S},
\end{eqnarray}
where the contribution from the angular part is symbolically denoted by
\begin{equation} \label{frakS}
\mathfrak{S} = 2 \pi \int d\theta \sin\theta \Theta \, \Theta^*,
\end{equation}
which will contribute the ``same'' factor for the flux at the near horizon region and at the asymptotic region. Therefore, it does not show up in the flux ratios, and consequently, it will not affect the physical quantities such as the mean number of produced pairs and the absorption cross section ratio, etc.

%%%%%%%%%%%%%%%%%%%%%%%%%%%%%%%%%%%%%%%%%%%%%%%%%%%%%%%%%%%%%%%%%%%%%%
\section{Pair Production}\label{pairproduction}
%%%%%%%%%%%%%%%%%%%%%%%%%%%%%%%%%%%%%%%%%%%%%%%%%%%%%%%%%%%%%%%%%%%%%%
The general solution of the radial equation~(\ref{EqR}) can be found in terms of the hypergeometric functions,
\begin{eqnarray} \label{sol}
R(r) &=& c_1 (r - B)^{\frac{i}2 (\tilde \kappa - \kappa)} (r + B)^{\frac{i}2 (\tilde \kappa + \kappa)} \, F\left( \frac12 + i \tilde \kappa + i \mu, \frac12 + i \tilde \kappa - i \mu; 1 + i \tilde \kappa - i \kappa; \frac12 - \frac{r}{2B} \right)
\nonumber\\
&+& c_2 (r - B)^{- \frac{i}2 (\tilde \kappa - \kappa)} (r + B)^{\frac{i}2 (\tilde \kappa + \kappa)} \, F\left( \frac12 + i \kappa + i \mu, \frac12 + i \kappa - i \mu; 1 - i \tilde \kappa + i \kappa; \frac12 - \frac{r}{2B} \right),
\end{eqnarray}
with parameters
\begin{equation} \label{kappamu}
\tilde \kappa = \frac{\omega}{B}, \qquad \kappa = \frac{q Q^3 - 2 n a r_0}{r_0^2 + a^2}, \qquad \mu = \sqrt{\kappa^2 - m^2 (r_0^2 + a^2) - \lambda - \frac14},
\end{equation}
in which $\mu^2$ is positive due to the BF bound violation in Eq.~(\ref{BFbv}). It turns out that this solution is a generalization of the corresponding result for the RN black holes~\cite{Chen:2012zn} with different parameters.

In the near horizon region, the solution~(\ref{sol}) can be expanded around $r = B$ (we assume $\tilde{\kappa} > \kappa$ to cover the extremal limit $B \to 0$)
\begin{equation}
R_\mathrm{H}(r) \approx c_\mathrm{H}^\mathrm{(in)} (2B)^{\frac{i}2 (\tilde \kappa + \kappa)} (r - B)^{- \frac{i}2 (\tilde \kappa - \kappa)} + c_\mathrm{H}^\mathrm{(out)} (2B)^{\frac{i}2 (\tilde \kappa + \kappa)} (r - B)^{\frac{i}2 (\tilde \kappa - \kappa)},
\end{equation}
where
\begin{equation}
c_\mathrm{H}^\mathrm{(in)} = c_2, \qquad c_\mathrm{H}^\mathrm{(out)} = c_1.
\end{equation}
In the asymptotic boundary ($r \to \infty$) of the metric (\ref{NHKN}), the hypergeometric function can be transformed into another form (see appendix~\ref{AHF}), so that we have
\begin{eqnarray} \label{RBasy}
R_\mathrm{B}(r) &\approx& c_\mathrm{B}^\mathrm{(in)} (r - B)^{-\frac12 - \frac{i}2 (\tilde \kappa + \kappa) - i \mu} (r + B)^{\frac{i}2 (\tilde \kappa + \kappa)} + c_\mathrm{B}^\mathrm{(out)} (r - B)^{-\frac12 - \frac{i}2 (\tilde \kappa + \kappa) + i \mu} (r + B)^{\frac{i}2 (\tilde \kappa + \kappa)}
\nonumber\\
&\approx& c_\mathrm{B}^\mathrm{(in)} r^{-\frac12 - i \mu} + c_\mathrm{B}^\mathrm{(out)} r^{-\frac12 + i \mu},
\end{eqnarray}
where
\begin{eqnarray} \label{cbin}
c_\mathrm{B}^\mathrm{(in)} &=& c_1 (2 B)^{\frac12 + i \tilde \kappa + i \mu} \frac{\Gamma(1 + i \tilde \kappa - i \kappa) \Gamma(- 2 i \mu)}{\Gamma\left( \frac12 - i \kappa - i \mu \right) \Gamma\left( \frac12 + i \tilde \kappa - i \mu \right)}
\nonumber\\
&& + c_2 (2 B)^{\frac12 + i \kappa + i \mu} \frac{\Gamma(1 - i \tilde \kappa + i \kappa) \Gamma(- 2 i \mu)}{\Gamma\left( \frac12 + i \kappa - i \mu \right) \Gamma\left( \frac12 - i \tilde \kappa - i \mu \right)},
\\ \label{cbout}
c_\mathrm{B}^\mathrm{(out)} &=& c_1 (2 B)^{\frac12 + i \tilde \kappa - i \mu} \frac{\Gamma(1 + i \tilde \kappa - i \kappa) \Gamma( 2 i \mu)}{\Gamma\left( \frac12 - i \kappa + i \mu \right) \Gamma\left( \frac12 + i \tilde \kappa + i \mu \right)}
\nonumber\\
&& + c_2 (2 B)^{\frac12 + i \kappa - i \mu} \frac{\Gamma(1 - i \tilde \kappa + i \kappa) \Gamma(2 i \mu)}{\Gamma\left( \frac12 + i \kappa + i \mu \right) \Gamma\left( \frac12 - i \tilde \kappa + i \mu \right)}.
\end{eqnarray}
Using the approximate solutions near the both inner and outer boundaries, the corresponding fluxes of each mode can be obtained via~(\ref{fluxD})
\begin{eqnarray}
D_\mathrm{B}^\mathrm{(in)} = - 2 (r_0^2 + a^2) \mu \bigl| c_\mathrm{B}^\mathrm{(in)} \bigr|^2 \mathfrak{S}, &\qquad&  D_\mathrm{H}^\mathrm{(in)} = - 2 B (r_0^2 + a^2) (\tilde \kappa - \kappa) \bigl| c_\mathrm{H}^\mathrm{(in)} \bigr|^2 \mathfrak{S},
\nonumber\\
D_\mathrm{B}^\mathrm{(out)} = 2 (r_0^2 + a^2) \mu \bigl| c_\mathrm{B}^\mathrm{(out)} \bigr|^2 \mathfrak{S}, &\qquad&  D_\mathrm{H}^\mathrm{(out)} = 2 B (r_0^2 + a^2) (\tilde \kappa - \kappa) \bigl| c_\mathrm{H}^\mathrm{(out)} \bigr|^2 \mathfrak{S},
\end{eqnarray}
where $\mathfrak{S}$ is defined in Eq.~(\ref{frakS}).

\begin{figure}
\centering
\includegraphics[width=4in]{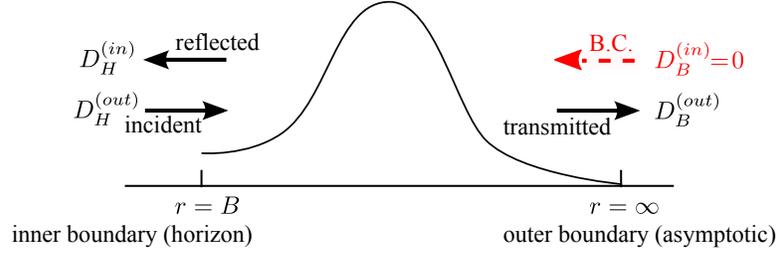}
\caption{Outer boundary condition: no incoming flux at asymptotic.}
\label{fig}
\end{figure}

For scalar particles, the flux conservation is $|D_\mathrm{incident}| = |D_\mathrm{reflected}| + |D_\mathrm{transmitted}|$, and the Bogoliubov relation is $|\alpha|^2 - |\beta|^2 = 1$, in which the vacuum persistence amplitude $|\alpha|^2$ and the mean number of produced pairs $|\beta|^2$ are given by the ratios of the fluxes in the Coulomb gauge
\begin{equation} \label{Bogoliubov}
|\alpha|^2 \equiv \frac{D_\mathrm{incident}}{D_\mathrm{reflected}}, \qquad |\beta|^2 \equiv \frac{D_\mathrm{transmitted}}{D_\mathrm{reflected}}.
\end{equation}
Moreover, from the viewpoint of scattering process of an incident flux from the asymptotic boundary, we can define the absorption cross section ratio as
\begin{equation} \label{absorptioncrosssection}
\sigma_\mathrm{abs} \equiv \frac{D_\mathrm{transmitted}}{D_\mathrm{incident}} = \frac{|\beta|^2}{|\alpha|^2}.
\end{equation}
The spontaneous pair production can be revealed by two different boundary conditions, namely, the outer boundary condition in which there is no incoming flux at the asymptotic outer boundary, and the inner boundary condition in which there is no outgoing flux at the black hole horizon. It was shown that these two boundary conditions are actually equivalent to each other due to the unitarity of the scattering matrix and give the same result~\cite{Chen:2012zn}. In the following, we will adopt the particle point of view, i.e., impose the outer boundary condition~(see Fig.~\ref{fig}) by setting $c_\mathrm{B}^\mathrm{(in)} = 0$. By substituting
\begin{equation}
c_1 = - c_2 (2B)^{i (\kappa - \tilde\kappa)} \frac{\Gamma(1 - i \tilde\kappa + i \kappa) \Gamma(\frac12 + i \tilde\kappa - i \mu) \Gamma(\frac12 - i \kappa - i \mu)}{\Gamma(\frac12 + i \kappa - i \mu) \Gamma(\frac12 - i \tilde\kappa - i \mu) \Gamma(1 - i \kappa + i \tilde\kappa)},
\end{equation}
into Eq.~(\ref{cbout}), we have
\begin{equation}
c_\mathrm{B}^\mathrm{(out)} = - c_2 (2B)^{\frac12 + i \kappa - i \mu} \frac{\sinh(2 \pi \mu) \sinh(\pi \tilde\kappa - \pi \kappa)}{\cosh(\pi \tilde\kappa - \pi \mu) \cosh(\pi \kappa + \pi \mu)} \frac{\Gamma(1 - i \tilde\kappa + i \kappa) \Gamma(2 i \mu)}{\Gamma(\frac12 - i \tilde\kappa + i \mu) \Gamma(\frac12 + i \kappa + i \mu)},
\end{equation}
and obtain the vacuum persistence amplitude and the mean number, respectively,
\begin{eqnarray} \label{Bogoliubovalpha}
|\alpha|^2 &=& \frac{D_\mathrm{incident}}{D_\mathrm{reflected}} = \frac{\bigl| D_\mathrm{H}^\mathrm{(out)} \bigr|}{\bigl| D_\mathrm{H}^\mathrm{(in)} \bigr|} = \frac{\cosh(\pi \kappa - \pi \mu) \cosh(\pi \tilde\kappa + \pi \mu)}{\cosh(\pi \kappa + \pi \mu) \cosh(\pi \tilde\kappa - \pi \mu)},
\\ \label{Bogoliubovbeta}
|\beta|^2 &=& \frac{D_\mathrm{transmitted}}{D_\mathrm{reflected}} = \frac{\bigl| D_\mathrm{B}^\mathrm{(out)} \bigr|}{\bigl| D_\mathrm{H}^\mathrm{(in)} \bigr|} = \frac{\sinh(2 \pi \mu) \sinh(\pi \tilde\kappa - \pi \kappa)}{\cosh(\pi \kappa + \pi \mu) \cosh(\pi \tilde\kappa - \pi \mu)},
\end{eqnarray}
and the absorption cross section ratio
\begin{equation} \label{Nsigma}
\sigma_\mathrm{abs} = \frac{D_\mathrm{transmitted}}{D_\mathrm{incident}} = \frac{\bigl| D_\mathrm{B}^\mathrm{(out)} \bigr|}{\bigl| D_\mathrm{H}^\mathrm{(out)} \bigr|} = \frac{\sinh(2 \pi \mu) \sinh(\pi \tilde \kappa - \pi \kappa)}{\cosh(\pi \kappa - \pi \mu) \cosh(\pi \tilde \kappa + \pi \mu)}.
\end{equation}
Similarly to the RN black hole case, $|\beta|^2$ is related with $\sigma_\mathrm{abs}$ through $|\beta|^2(\mu \rightarrow -\mu) \rightarrow -\sigma_\mathrm{abs}$. If we take the extremal limit $B \to 0$ i.e. $\tilde \kappa \to \infty$, we have
\begin{equation} \label{ex vac-per}
|\alpha|^2 = \frac{\cosh(\pi \kappa - \pi \mu)}{\cosh(\pi \kappa + \pi \mu)} e^{2 \pi \mu}, \qquad |\beta|^2 = \frac{\sinh(2 \pi \mu)}{\cosh(\pi \kappa + \pi \mu)} e^{\pi \mu - \pi \kappa}, \qquad \sigma_\mathrm{abs} = \frac{\sinh(2 \pi \mu)}{\cosh(\pi \kappa - \pi \mu)} e^{-\pi \mu - \pi \kappa}.
\end{equation}

%\subsection{Angular Part}
%%%%%%%%%%%%%%%%%%%%%%%%%%%%%%%%%%%%%%%%%%%%%%%%%%%%%%%%%%%%%%%%%%%%%%
For the sake of self-completeness, the angular equation for $\Theta(\theta)$ can be expressed as
\begin{equation} \label{AEq}
\frac1{\sin\theta} \partial_\theta (\sin\theta \partial_\theta \Theta) - \left( \frac{n^2}{\sin^2\theta} - \lambda_1 \sin^2\theta - \lambda_2 \right) \Theta = 0,
\end{equation}
where
\begin{equation}
\lambda_1 = m^2 a^2 - \frac{(q Q r_0 - n a)^2}{(r_0^2 + a^2)^2} a^2, \qquad \lambda_2 = \lambda - \frac{2 n a (q Q r_0 - n a)}{r_0^2 + a^2}.
\end{equation}
This equation can be transformed into the confluent Heun equation. But, the angular part contributes the ``same'' factor for the flux both at the near horizon region and at the asymptotic region. Therefore, its explicit solution is not necessary for our consideration of pair production. However, the regularity for the angular solution should give constraint on the separation parameter $\lambda$, which can be found numerically. Another interesting remark is that the existence condition for the pair production (violation of the BF bound) for RN black holes just turns out to guarantee the cosmic censorship conjecture during the pair production process. Such a connection, however, is not so obvious for KN black holes.

\section{Thermal interpretation}\label{thermalinter}
%%%%%%%%%%%%%%%%%%%%%%%%%%%%%%%%%%%%%%%%%%%%%%%%%%%%%%%%%%%%%%%%%%%%%%
In terms of instanton actions $S_a = 2 \pi \kappa, \tilde{S}_a = 2 \pi \tilde{\kappa}$ and $S_b = 2 \pi \mu$ from the relativistic field equation, we write the mean number of produced pairs~(\ref{Bogoliubovbeta}) as
\begin{equation}
%\label{vacuumpersistenceamplitude}
%|\alpha|^2 &=& \left(\frac{1 + \mathrm{e}^{-S_a + S_b}}{1 + \mathrm{e}^{-S_a - S_b}} \right)
%\left( \frac{1 + \mathrm{e}^{-\tilde{S}_a - S_b}}{1 + \mathrm{e}^{-\tilde{S}_a + S_b}} \right), \\
\label{meannumberpairs}
\mathcal{N} = |\beta|^2 = \left( \frac{\mathrm{e}^{-S_a + S_b} - \mathrm{e}^{-S_a - S_b}}{1 + \mathrm{e}^{-S_a - S_b}} \right) \left( \frac{1 - \mathrm{e}^{-\tilde{S}_a + S_a}}{1 + \mathrm{e}^{-\tilde{S}_a + S_b}} \right).
\end{equation}
Note that the mean number (\ref{meannumberpairs}) has a similar form except for different quantum numbers as that of charged scalars in a near-extremal RN black hole in Ref.~\cite{Chen:2012zn} since the near-horizon geometry is the warped ${\rm AdS}_3 \times S^1$ for the near-extremal KN black hole while it is the ${\rm AdS}_2 \times S^2$ for the  near-extremal RN black hole. Following Refs.~\cite{Kim2015a, Kim2015b}, by introducing an effective temperature and its associated one
\begin{equation} \label{effectivetemperature}
T_\mathrm{KN} = \frac{\bar{m}}{S_a - S_b} = T_U + \sqrt{T_U^2 + \frac{\mathcal{R}}{8 \pi^2}}, \qquad \bar{T}_\mathrm{KN} = \frac{\bar{m}}{S_a + S_b} = T_U - \sqrt{T_U^2 + \frac{\mathcal{R}}{8 \pi^2}},
\end{equation}
where
\begin{equation}
\bar{m} = \sqrt{m^2 - \frac{\lambda + 1/4}{2} \mathcal{R}},
\end{equation}
and
\begin{equation} \label{unruh}
T_U = \frac{\kappa}{2 \pi \bar{m}(r_0^2 + a^2)} = \frac{q Q^3 - 2 n a r_0}{2 \pi \bar{m}(r_0^2 + a^2)^2}, \qquad \mathcal{R} = - \frac{2}{r_0^2 + a^2},
\end{equation}
the mean number (\ref{meannumberpairs}) can be expressed as
\begin{equation} \label{thermal}
\mathcal{N} = \mathrm{e}^{\frac{\bar{m}}{T_\mathrm{KN}}} \times \left( \frac{\mathrm{e}^{-\frac{\bar{m}}{T_\mathrm{KN}}} - \mathrm{e}^{-\frac{\bar{m}}{\bar{T}_\mathrm{KN}}}}{1 + \mathrm{e}^{-\frac{\bar{m}}{\bar{T}_\mathrm{KN}}}} \right) \times \left\{ \frac{\mathrm{e}^{-\frac{\bar{m}}{T_\mathrm{KN}}} \bigl( 1 - \mathrm{e}^{-\frac{\hat{\omega} - q \Phi_\mathrm{H} - n \Omega_\mathrm{H}}{T_\mathrm{H}}} \bigr)}{1 + \mathrm{e}^{-\frac{\hat{\omega} - q \Phi_\mathrm{H} - n \Omega_\mathrm{H}}{T_\mathrm{H}}} \mathrm{e}^{- \frac{\bar{m}}{T_\mathrm{KN}}}} \right\}.
\end{equation}
Here, $T_\mathrm{H}$ is the Hawking temperature and the $\Phi_\mathrm{H}, \Omega_\mathrm{H}$ are the chemical potential and angular velocity at $r = B$
\begin{equation}
T_\mathrm{H} = \frac{\hat{B}}{2 \pi}, \qquad \Phi_\mathrm{H} = - \frac{Q^3 \hat{B}}{r_0^2 + a^2}, \qquad \Omega_\mathrm{H} = \frac{2 a r_0 \hat{B}}{r_0^2 + a^2},
\end{equation}
in which $\hat{\omega} = \varepsilon \omega$ and $\hat{B} = \varepsilon B$ are quantities measured in the ``original'' coordinates of KN black holes (\ref{KN}). However, the result (\ref{thermal}) is independent on the parameter $\varepsilon$, thus here and after we do not elaborately distinguish $(\hat \omega, \hat B)$ from $(\omega, B)$ in such a situation.

The physical interpretation of each term of Eq.~(\ref{thermal}) is that the first parenthesis is the Schwinger effect with the effective temperature $T_\mathrm{KN}$ in ${\rm AdS}_2$~\cite{Cai2014} and the second parenthesis is the Schwinger effect in the Rindler space~\cite{Gabriel2000}, in which the Unruh temperature is given by the Hawking temperature and the charge has the chemical potentials of $\Phi_\mathrm{H}$ and $\Omega_\mathrm{H}$ while the effective temperature for the Schwinger effect due to the electric field on the horizon is determined by $T_\mathrm{KN}$.

The mean number of produced pairs above and the absorption cross section ratio in the previous section have been obtained using the exact solution in the near horizon geometry of an extremal or near extremal KN black hole. Below, by applying the phase-integral formula, we derive the instanton actions from the Hamilton-Jacobi action for the field equation, which lead to the mean number. This method allows one to understand the physical origin of each term as a consequence of simple poles in the complex plane of space~\cite{Kim2007} and further also connects the interpretation to other physical systems involving the Schwinger effect in curved spacetimes~\cite{Kim2013}. Also the acceleration of the scalar charge near the horizon gives the Unruh temperature when evaluated on the horizon.

\subsection{Hamilton-Jacobi Action and Phase-Integral Formula} 
%%%%%%%%%%%%%%%%%%%%%%%%%%%%%%%%%%%%%%%%%%%%%%%%%%%%%%%%%%%%%%%%%%%%%%
The Hamilton-Jacobi action together with the phase-integral formula explains the origin of the subleading terms as well as the leading Boltzmann factor in Eq.~(\ref{meannumberpairs}). By substituting $R(r) = \mathrm{e}^{i S(r)}$ into the radial equation~(\ref{EqR}), we obtain the Hamilton-Jacobi action in the complex plane of $z=r$
\begin{equation} \label{integral}
S(z) = \int \frac{dz}{z^2 - B^2} \sqrt{(\omega - \kappa z)^2 - \bar{m}^2 (r_0^2 + a^2)(z^2 - B^2)}.
\end{equation}
The phase-integral formula for the action along a path $\Gamma$ in the $z$-plane gives a complex amplitude
\begin{equation} \label{phaseintegral}
N_S = \mathrm{e}^{i S_\Gamma},
\end{equation}
whose imaginary part due to simple poles determines the pair production rate or the decay rate of the vacuum \cite{Kim2007,Chen:2012zn}.

\begin{figure}
\centering
\includegraphics[width=3in]{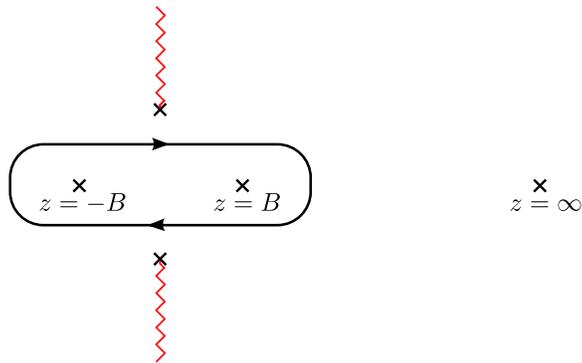}
\caption{The contour integral for the leading Boltzmann factor in the extremal KN black hole.}
\label{fig2}
\end{figure}

\begin{figure}
\centering
\includegraphics[width=3in]{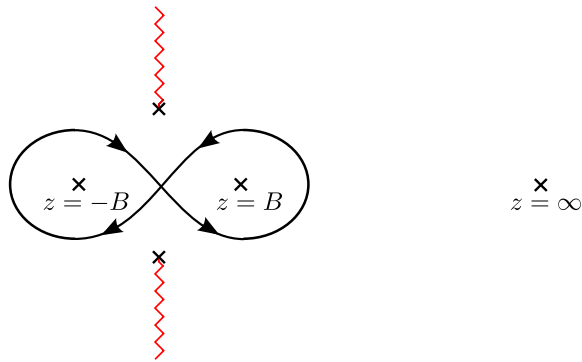}
\caption{The contour integral for the sub-leading term in the near extremal KN black hole.}
\label{fig3}
\end{figure}

First, in the case of the near extremal KN black hole, the action (\ref{integral}) has two simple poles at $z = \pm B$ and another simple pole at the infinity $z = \infty$ from the large $z$-expansion and thereby the residues $S_{\pm}$ and $S_{\infty}$, respectively. Further, the square root has a pair of complex roots in the complex plane, so a branch cut may be introduced as in Fig.~\ref{fig2} to exclude the roots and make the integrand an analytic function. Then, the Cauchy residue theorem along the contour Fig.~\ref{fig2} enclosing the simple poles at $z = \pm B$ gives the leading term
\begin{eqnarray}\label{phaseintegral1}
\mathrm{e}^{i (-2 \pi i)( \frac{\omega - \kappa B}{2B} - \frac{\omega + \kappa B}{2B} + \mu)} = \mathrm{e}^{- (S_a - S_{\infty})} = \mathrm{e}^{- \frac{\bar{m}}{T_\mathrm{KN}}},
\end{eqnarray}
where the third term in the parenthesis is the residue from the infinity. On the other hand, the contour integral of Fig.~\ref{fig3} provides another instanton action for the exact pair production rate (\ref{meannumberpairs})
\begin{eqnarray}
\tilde{S}_a = (S_- - S_{\infty}) - (S_+ - S_{\infty}) = 2 \pi \frac{\omega}{B} = 2 \pi \tilde{\kappa}.
\end{eqnarray}
A few comments are in order. The directions of the contours are chosen by the causality reason. The Schwinger effect in a near extremal black hole is a consequence of both the inner and outer horizons, which differs from the Hawking radiation of charges in a non-extremal black hole originated from the outer horizon only.

Second, in the case of the extremal KN black hole, we use the small $z$-expansion, which has a simple pole at $z=0$ and the residue $- \kappa$. Thus, the clockwise contour integral gives the leading term
\begin{eqnarray} \label{ex phaseintegral1}
\mathrm{e}^{i (-2 \pi i)( - \kappa + \mu)} = \mathrm{e}^{- (S_a - S_{\infty})} = \mathrm{e}^{- \frac{\bar{m}}{T_\mathrm{KN}}},
\end{eqnarray}
where the second term comes from the simple pole at the infinity. Note that Eq. (\ref{ex phaseintegral1}) is the same as Eq. (\ref{phaseintegral1}).

\subsection{Unruh temperature}
%%%%%%%%%%%%%%%%%%%%%%%%%%%%%%%%%%%%%%%%%%%%%%%%%%%%%%%%%%%%%%%%%%%%%%
We now explain the origin of the Unruh temperature (\ref{unruh}) for accelerating charges. For that purpose, we use the two conserved quantities associated with two killing vectors of the background~(\ref{NHKN}), $\partial_t$ and $\partial_\varphi$, which are given by
\begin{eqnarray}
E &=& \left[ \Gamma(\theta)(r^2 - B^2) - \gamma(\theta) b^2 r^2 \right] u^t - \gamma(\theta) b r u^\varphi,
\\
L &=& \gamma(\theta) (u^\varphi + b r u^t).
\end{eqnarray}
We consider a trajectory of a charged particle without carrying angular momentum, $L = 0$ such that $u^\varphi = - b r u^t$, and also $u^\theta = 0$. We can then express the 4-velocity with $-1 = g_{\mu\nu} u^\mu u^\nu$ in terms of the energy of moving particle as
\begin{eqnarray} \label{velocity}
u^t &=& \frac{E}{\Gamma(\theta)(r^2 - B^2)},
\nonumber\\
u^r &=& \frac{\sqrt{E^2 - (r^2 - B^2) \Gamma(\theta)}}{\Gamma(\theta)},
\nonumber\\
u^\varphi &=& \frac{-2 E a r_0 r}{\Gamma(\theta) (r^2 - B^2) (r_0^2 + a^2)}.
\end{eqnarray}
From the equation of motion for the charge in a curved spacetime under an electromagnetic force
\begin{equation} \label{geodesic}
\frac{du^\mu}{ds} + \Gamma^\mu{}_{\alpha \beta} u^\alpha u^\beta = \frac{q}{m} F^\mu{}_\nu u^\nu,
\end{equation}
we may compute the corresponding acceleration by
\begin{eqnarray} \label{acceleration}
|a^\mu| = \left| \frac{du^\mu}{ds} \right| = \sqrt{\frac{q^2}{m^2} F^\mu{}_\alpha F_{\mu \beta} u^\alpha u^\beta + \Gamma_{\mu \alpha \beta} \Gamma^\mu{}_{\gamma \delta} u^\alpha u^\beta u^\gamma u^\delta},
\end{eqnarray}
for the particular case $u^t = \pm \sqrt3 \, u^r/(r^2 - B^2)$ leading to have vanishing the crossing term $F_{\mu \alpha}\Gamma^\mu{}_{\beta \gamma} u^\alpha u^\beta u^\gamma = 0$. Thus, by substituting
\begin{equation}
E = \pm \sqrt{\frac32 (r^2 - B^2) \Gamma(\theta)},
\end{equation}
into~(\ref{velocity}) and~(\ref{acceleration}), we have
\begin{equation}
|a^\mu| = \sqrt{\frac{q^2 Q^2 (r_0^2 - a^2 \cos^2\theta)^2}{m^2 \Gamma(\theta)^4} - \frac{2 r^2}{(r^2 - B^2) \Gamma(\theta)} + \frac{a^2 \sin^2\theta (3 r_0^2 + a^2 \cos^2\theta)}{\Gamma(\theta)^3}}.
\end{equation}
On the polar points $(\theta = 0, \pi)$, where thermodynamical variables are computed, and in the extremal limit $(B = 0)$, the acceleration reduces to
\begin{equation}
|a^\mu| = \sqrt{\frac{q^2 Q^6}{m^2 (r_0^2 + a^2)^4} - \frac2{r_0^2 + a^2}},
\end{equation}
and the corresponding Unruh temperature is that of (\ref{unruh}) with zero quantum numbers $(n = 0, \lambda + 1/4 =0)$
\begin{equation}
T_U = \frac{|a^\mu|}{2 \pi} = \sqrt{\frac{q^2 Q^6}{4 \pi^2 m^2 (r_0^2 + a^2)^4} - \frac1{2 \pi^2 (r_0^2 + a^2)}}.
\end{equation}
Note that the worldline instanton for the Schwinger effect~\cite{Dunne2005} from the trajectory (\ref{geodesic}) does not carry quantum numbers or transverse fluctuations whereas the instanton action in previous subsection, which carries all quantum numbers for fluctuations and includes the prefactor~\cite{Kim2007}. Thus, the acceleration of the charge explains the Unruh temperature for the Schwinger effect.

%%%%%%%%%%%%%%%%%%%%%%%%%%%%%%%%%%%%%%%%%%%%%%%%%%%%%%%%%%%%%%%%%%%%%%
\section{Dual CFT Description}\label{cftdescription}
%%%%%%%%%%%%%%%%%%%%%%%%%%%%%%%%%%%%%%%%%%%%%%%%%%%%%%%%%%%%%%%%%%%%%%
The KN black hole was shown to have twofold dual CFT descriptions~\cite{Chen:2010ywa, Chen:2012np} which are called the $J$ (angular momentum) picture in terms of the Kerr/CFT correspondence~\cite{Guica:2008mu, Compere:2012jk} and the $Q$ (charge) picture based on the RN/CFT correspondence~\cite{Chen:2009ht, Chen:2011gz}. Following previous studies on the holographic description of the pair production in the near extremal RN black hole~\cite{Chen:2015fkx}, it is interesting to analyze the holographic dual of the pair production in the KN black hole in both of the two pictures.

The absorption cross section ratio of the charged scalar field in eq.~(\ref{Nsigma}) can be rewritten as
\begin{equation} \label{absorption}
\sigma_\mathrm{abs} = \frac{\sinh(2 \pi \mu)}{\pi^2} \sinh(\pi \tilde\kappa - \pi \kappa) \left| \Gamma\left( \frac12 + i (\mu - \kappa) \right) \right|^2 \left| \Gamma\left( \frac12 + i (\mu + \tilde\kappa) \right) \right|^2,
\end{equation}
with the parameters $\tilde\kappa, \kappa$, and $\mu$ defined in~(\ref{kappamu}). From the field/operator duality, the charged scalar field in the near extremal KN black hole is dual to a scalar operator in the two-dimensional CFT. In the present case, the conformal dimensions $(h_\mathrm{L}, h_\mathrm{R})$ of the operator can be read from the asymptotic expansion of the bulk charged scalar field at the AdS boundary eq.~(\ref{RBasy}), which are given by~\cite{Bredberg:2009pv}
\begin{equation}
h_\mathrm{L} = h_\mathrm{R} = \frac12 + i \mu.
\end{equation}
On the other hand, the absorption cross section ratio of the dual scalar operator in the two-dimensional CFT has the universal form, namely
\begin{equation} \label{CFTabs}
\sigma_\mathrm{abs} \sim T_\mathrm{L}^{2 h_\mathrm{L} - 1} T_\mathrm{R}^{2 h_\mathrm{R} - 1} \sinh\left( \frac{{\tilde\omega}_\mathrm{L}}{2 T_\mathrm{L}} + \frac{{\tilde\omega}_\mathrm{R}}{2 T_\mathrm{R}} \right) \left| \Gamma\left( h_\mathrm{L} + i \frac{{\tilde\omega}_\mathrm{L}}{2 \pi T_\mathrm{L}} \right) \right|^2 \left| \Gamma\left( h_\mathrm{R} + i \frac{{\tilde\omega}_\mathrm{R}}{2 \pi T_\mathrm{R}} \right) \right|^2,
\end{equation}
where $T_\mathrm{L}$ and $T_\mathrm{R}$ are left- and right-hand temperature of the dual CFT, ${\tilde\omega}_\mathrm{L} = \omega_\mathrm{L} - q_\mathrm{L} \Phi_\mathrm{L}$ and ${\tilde\omega}_\mathrm{R} = \omega_\mathrm{R} - q_\mathrm{R} \Phi_\mathrm{R}$ are the total excited energy of the left- and right-hand sectors in which $(q_\mathrm{L}, q_\mathrm{R})$ and $(\Phi_\mathrm{L}, \Phi_\mathrm{R})$ are respectively the charges and chemical potential of the dual left and right-hand operators.

Before comparing eq.~(\ref{absorption}) with eq.~(\ref{CFTabs}), let us recall that for the near extremal KN black hole, its entropy and temperature are
\begin{eqnarray}
&& S_\mathrm{BH} = \pi (\hat{r}_+^2 + a^2) \quad \Rightarrow \quad S_\mathrm{BH} \sim \pi (r_0^2 + a^2 + 2 r_0 B),
\\
&& T_\mathrm{H} = \frac{\hat{r}_+ - \hat{r}_-}{4 \pi (\hat{r}_+^2 + a^2)} \quad \Rightarrow \quad T_\mathrm{H} = \frac{B}{2 \pi}.
\end{eqnarray}
Then we will discuss the holographic description of the pair production in the near extremal KN black hole in the $J$ and $Q$ pictures, respectively.

\subsection{$J$-picture}
%%%%%%%%%%%%%%%%%%%%%%%%%%%%%%%%%%%%%%%%%%%%%%%%%%%%%%%%%%%%%%%%%%%%%%
The left- and right-hand central charges for the $J$-picture are~\cite{Chen:2010ywa, Chen:2012np}
\begin{equation}
c_\mathrm{L}^J = c_\mathrm{R}^J = 12 J,
\end{equation}
and the associated temperatures are
\begin{equation}
T_\mathrm{L}^J = \frac{\hat{r}_+^2 + \hat{r}_-^2 + 2 a^2}{4 \pi a (\hat{r}_+ + \hat{r}_-)}, \qquad T_\mathrm{R}^J = \frac{\hat{r}_+ - \hat{r}_-}{4 \pi a},
\end{equation}
leads to the near extreme results
\begin{equation}
T_\mathrm{L}^J = \frac{r_0^2 + a^2}{4 \pi a r_0}, \qquad T_\mathrm{R}^J = \frac{B}{2 \pi a}.
\end{equation}
The CFT microscopic entropy is calculated from the Cardy formula via
\begin{equation}
S_\mathrm{CFT} = \frac{\pi^2}3 (c_\mathrm{L} T_\mathrm{L} + c_\mathrm{R} T_\mathrm{R}) = \pi (r_0^2 + a^2 + 2 r_0 B),
\end{equation}
which agrees with area entropy of the near extremal KN black hole.

Furthermore, we need to identify the first law of thermodynamics of the black hole with that of the dual CFT, i.e., $\delta S_\mathrm{BH} = \delta S_\mathrm{CFT}$
\begin{equation}\label{1stlaws}
\frac{\delta M - \Omega_\mathrm{H} \delta J - \Phi_\mathrm{H} \delta Q}{T_\mathrm{H}} = \frac{\tilde{\omega}_\mathrm{L}}{T_\mathrm{L}} + \frac{\tilde{\omega}_\mathrm{R}}{T_\mathrm{R}},
\end{equation}
in which the angular velocity and chemical potential (at $r = B$) are
\begin{equation}
\Omega_\mathrm{H} = \frac{2 a r_0}{r_0^2 + a^2} B, \qquad \Phi_\mathrm{H} = - \frac{Q^3 B}{r_0^2 + a^2}.
\end{equation}
Note that in the $J$ picture, $T_\mathrm{L} = T_\mathrm{L}^J$ and $T_\mathrm{R} = T_\mathrm{R}^J$. In addition, the charge of the probe scalar field is turned off in order to only ``feel'' the rotation of the KN black hole, thus the variation of the conserved charges of the KN black hole are $\delta M = \omega, \; \delta J = -n, \; \delta Q = 0$. Subsequently, the excitation of the total energy for the dual $J$ picture CFT are determined as
\begin{equation}
{\tilde \omega}_\mathrm{L}^J = n \quad \mathrm{and} \quad {\tilde \omega}_\mathrm{R}^J = \frac{\omega}{a},
\end{equation}
then we have $\frac{{\tilde\omega}_\mathrm{L}}{2 T_\mathrm{L}} = - \pi \kappa$ and $\frac{{\tilde\omega}_\mathrm{R}}{2 T_\mathrm{R}} = \pi \tilde\kappa$ in which $q$ is set to zero. Consequently, in the $J$ picture of the KN/CFTs duality, the absorption cross section ratio of the scalar field (with $q = 0$) in eq.~(\ref{absorption}) matches well with that of its dual scalar operator in eq.~(\ref{CFTabs}).

\subsection{$Q$-picture}
%%%%%%%%%%%%%%%%%%%%%%%%%%%%%%%%%%%%%%%%%%%%%%%%%%%%%%%%%%%%%%%%%%%%%%
The left- and right-hand central charges for the $Q$-picture are~\cite{Chen:2010ywa, Chen:2012np}
\begin{equation}
c_\mathrm{L}^Q = c_\mathrm{R}^Q = \frac{6 Q^3}{\ell},
\end{equation}
where the parameter $\ell$ is the measure of the $U(1)$ bundle formed by the background Maxwell field, which has a geometrical interpretation of the radius of the embedded extra circle in the fifth dimension, and the associated left- and right-hand temperatures are
\begin{equation}
T_\mathrm{L}^Q = \frac{(\hat{r}_+^2 + \hat{r}_-^2 + 2 a^2) \ell}{4 \pi Q (\hat{r}_+ \hat{r}_- - a^2)}, \qquad T_\mathrm{R}^Q = \frac{(\hat{r}_+^2 - \hat{r}_-^2) \ell}{4 \pi Q (\hat{r}_+ \hat{r}_- - a^2)},
\end{equation}
which, in the near extreme case, can be expressed as
\begin{equation}
T_\mathrm{L}^Q = \frac{(r_0^2 + a^2) \ell}{2 \pi Q^3}, \qquad T_\mathrm{R}^Q = \frac{r_0 B \ell}{\pi Q^3}.
\end{equation}
Besides, the CFT microscopic entropy computed from the Cardy formula
\begin{equation}
S_\mathrm{CFT} = \frac{\pi^2}3 (c_\mathrm{L} T_\mathrm{L} + c_\mathrm{R} T_\mathrm{R}) = \pi (r_0^2 + a^2 + 2 r_0 B),
\end{equation}
reproduces the macroscopic entropy of the near extremal KN black hole, too.

In the $Q$ picture, the modes characterizing the rotation of the charged probe scalar field, namely $n$, should be turned off in order that the charged probe only interact with the charge of the black hole, i.e., $\delta M = \omega, \; \delta J = 0, \; \delta Q = -q$. Again, using the identification between the thermodynamics of the black hole and its dual CFT, namely, eq.~(\ref{1stlaws}), with $T_\mathrm{L} = T_\mathrm{L}^Q$ and $T_\mathrm{R} = T_\mathrm{R}^Q$. Then the excitation of the total energy of the dual $Q$ picture CFT are obtained as
\begin{equation}
{\tilde \omega}_\mathrm{L}^Q = -q \ell \quad \mathrm{and} \quad {\tilde \omega}_\mathrm{R}^Q = \frac{2 r_0 \ell \omega}{Q^3}.
\end{equation}
namely, $\frac{{\tilde\omega}_\mathrm{L}}{2 T_\mathrm{L}} = - \pi \kappa$ and $\frac{{\tilde\omega}_\mathrm{R}}{2 T_\mathrm{R}} = \pi \tilde\kappa$. Therefore, in the $Q$ picture of the KN/CFTs duality, the absorption cross section ratio of the charged scalar field (with $n = 0$) in eq.~(\ref{absorption}) also matches well with that of its dual scalar operator in eq.~(\ref{CFTabs}). Moreover, since the mean number of produced pairs $|\beta |^2$ of the charged probe scalar field is associated with its absorption cross section ratio via $|\beta|^2(\mu \to -\mu) \to -\sigma_\mathrm{abs}$, thus the holographic description of $|\beta|^2$ can be understood in both of the $J$- and $Q$-pictures.

%%%%%%%%%%%%%%%%%%%%%%%%%%%%%%%%%%%%%%%%%%%%%%%%%%%%%%%%%%%%%%%%%%%%%%
\section{Conclusions}\label{conclusion}
%%%%%%%%%%%%%%%%%%%%%%%%%%%%%%%%%%%%%%%%%%%%%%%%%%%%%%%%%%%%%%%%%%%%%%
In this paper, we have extended the previous study on the spontaneous production of charged pairs in the near extremal RN black to the case of the near extremal KN black hole. The near horizon warped AdS$_3$ geometry of the near extremal KN black hole allows us to obtain the analytic form of the vacuum persistence amplitude, the mean number of produced pairs and the absorption cross section ratio of charged scalars. We have found a universal feature of the Schwinger mechanism in extremal charged black holes, non-rotating or rotating. The violation of the BF bound of the charged scalar in the AdS$_2$ space is a necessary condition for the pair production. But its connection to the cosmic censorship conjecture is not so obvious as in the RN black holes.

The pair-production rate has a thermal interpretation with the effective temperature consisting of the Unruh temperature for accelerating charge and the ${\rm AdS}$ curvature. In fact, the Unruh temperature is the acceleration of charges at the polar points, though the exact formula predicts the Unruh temperature for charges carrying quantum numbers from the field equation. We have employed the Hamilton-Jacobi action and the phase-integral formula to compute the decay rate of the field due to the spontaneous pair production. It is shown that the Boltzmann factor for the Schwinger effect in the near extremal KN black hole comes from two simple poles located at the inner and outer horizons. There is no a priori reason to prevent the quantum tunneling process from the inner horizon since it is located just adjacent to the outer horizon contrary to a non-extremal KN black hole.

Since the near extremal KN black hole has the twofold dual CFT descriptions associated with the $J$ picture and the $Q$ picture, we have thus analyzed the holographic dual CFT descriptions of the pair production by comparing the absorption cross section ratios of the bulk charged scalar field and that of its dual scalar operator in the dual CFT in both of these two pictures and found they agree with each other. Furthermore, to present a clear description of the spontaneous pair production of scalars, our results also give more information about the CFT dual of the near extremal KN black hole. It would be interesting to take into account the back reaction to the background geometry to give a more precise picture of the pair production process.

\section*{Acknowledgments}
The authors would like to thank Pauchy Hwang for the warm hospitality at National Taiwan University, where part of this work was done, and also Rong-Gen Cai for useful discussions at 2016 Annual Meeting of the Chinese Society of Gravitation and Relativistic Astrophysics at Changsha. S.P.K. also would like to thank Robert Mann and Don N. Page for useful discussions during the BIRS-CMO Workshop on Black Holes' New Horizon and Fiorenzo Bastianelli for the warm hospitality and useful discussions at Bologna University. The work of C.M.C. was supported by the Ministry of Science and Technology of Taiwan under the grant MOST 102-2112-M-008-015-MY3. The work of S.P.K. was supported by Basic Science Research Program through the National Research Foundation of Korea (NRF) funded by the Ministry of Education (15B15770630). J.R.S. was supported by the National Natural Science Foundation of China (No.~11205058), the Open Project Program of State Key Laboratory of Theoretical Physics, Institute of Theoretical Physics, Chinese Academy of Sciences, China (No.~Y5KF161CJ1) and the Fundamental Research Funds for the Central Universities.

\begin{appendix}

\section{Hypergeometric Functions} \label{AHF}

There are a number of mathematical properties for the hypergeometric function $F(\alpha, \beta; \gamma; z)$, in particular the following one are useful for our analysis:
(i) transformation formula
\begin{eqnarray}\label{transformationformula}
F(\alpha, \beta; \gamma; z)
%&=& (1 - z)^{\gamma - \alpha - \beta} F(\gamma - \alpha, \gamma - \beta; \gamma; z)
%\\
%&=& (1 - z)^{-\alpha} F\left(\alpha, \gamma - \beta; \gamma; \frac{z}{z-1} \right) = (1 - z)^{-\beta} F\left( \beta, \gamma - \alpha; \gamma; \frac{z}{z-1} \right),
%\\
%&=& \frac{\Gamma(\gamma) \Gamma(\gamma - \alpha - \beta)}{\Gamma(\gamma - \alpha) \Gamma(\gamma - \beta)} F(\alpha, \beta; \alpha + \beta - \gamma + 1; 1 - z)
%\nonumber\\
%&& + \frac{\Gamma(\gamma) \Gamma(\alpha + \beta - \gamma)}{\Gamma(\alpha) \Gamma(\beta)} (1 - z)^{\gamma - \alpha - \beta} F(\gamma - \alpha, \gamma - \beta; \gamma - \alpha - \beta + 1; 1 - z),
%\\
%&=& \frac{\Gamma(\gamma) \Gamma(\beta - \alpha)}{\Gamma(\beta) \Gamma(\gamma - \alpha)} (-z)^{-\alpha} F\left( \alpha, \alpha - \gamma + 1; \alpha - \beta + 1; \frac1{z} \right)
%\nonumber\\
%&& + \frac{\Gamma(\gamma) \Gamma(\alpha - \beta)}{\Gamma(\alpha) \Gamma(\gamma - \beta)} (-z)^{-\beta} F\left( \beta, \beta - \gamma + 1; \beta - \alpha + 1; \frac1{z} \right),
%\\
&=& \frac{\Gamma(\gamma) \Gamma(\beta - \alpha)}{\Gamma(\beta) \Gamma(\gamma - \alpha)} (1 - z)^{-\alpha} F\left( \alpha, \gamma - \beta; \alpha - \beta + 1; \frac1{1 - z} \right)
\nonumber\\
&& + \frac{\Gamma(\gamma) \Gamma(\alpha - \beta)}{\Gamma(\alpha) \Gamma(\gamma - \beta)} (1 - z)^{-\beta} F\left( \beta, \gamma - \alpha; \beta - \alpha + 1; \frac1{1 - z} \right),
%\\
%&=& \frac{\Gamma(\gamma) \Gamma(\gamma - \alpha - \beta)}{\Gamma(\gamma - \alpha) \Gamma(\gamma - \beta)} z^{-\alpha} F\left( \alpha, \alpha - \gamma + 1; \alpha + \beta - \gamma + 1; 1 - \frac1{z} \right)
%\\
%&& + \frac{\Gamma(\gamma) \Gamma(\alpha + \beta - \gamma)}{\Gamma(\alpha) \Gamma(\beta)} (1 - z)^{\gamma - \alpha - \beta} z^{\alpha - \gamma} F\left( \gamma - \alpha, 1 - \alpha; \gamma - \alpha - \beta + 1; 1 - \frac1{z} \right), \label{T1} \nonumber
\end{eqnarray}
(ii) special values
\begin{equation}
F(\alpha, \beta; \gamma; 0) = 1, \qquad F(\alpha, \beta; \gamma; 1) = \frac{\Gamma(\gamma) \Gamma(\gamma - \alpha - \beta)}{\Gamma(\gamma - \alpha) \Gamma(\gamma - \beta)}.
\end{equation}
%and (iii) differential formula
%\begin{eqnarray}
%&& \partial_z F(\alpha, \beta; \gamma; z) = \frac{\alpha \beta}{\gamma} F(\alpha + 1, \beta + 1; %\gamma + 1; z),
%\\
%&& \partial_z \left[ z^\alpha F(\alpha, \beta; \gamma; z) \right] = \alpha z^{\alpha - 1} F(\alpha + 1, \beta; \gamma; z).
%\\
%&& \partial_z \left[ z^{\gamma -1} F(\alpha, \beta; \gamma; z) \right] = (\gamma - 1) z^{\gamma - 2} F(\alpha, \beta; \gamma - 1; z).
%\end{eqnarray}
In addition, the following properties of Gamma function are also needed in our computation
\begin{eqnarray}
&& \Gamma(\alpha + 1) = \alpha \Gamma(\alpha), \qquad \Gamma(\alpha) \Gamma(1 - \alpha) = \frac{\pi}{\sin(\alpha \pi)},
\\
&& \left| \Gamma\left( \frac12 + i y \right) \right|^2 = \frac{\pi}{\cosh(\pi y)}, \quad \left| \Gamma\left( 1 + i y \right) \right|^2 = \frac{\pi y}{\sinh(\pi y)}, \quad \left| \Gamma\left( i y \right) \right|^2 = \frac{\pi}{y \sinh(\pi y)}.
\end{eqnarray}

\end{appendix}

%%%%%%%%%%%%%%%%%%%%%%%%%%%%%%%%%%%%%%%%%%%%%%%%%%%%%%%%%%%%%%%%%%%%%%

\end{document}